\documentstyle[amssymb,multicol,epsf,aps]{revtex}
\def\dbarrm {{\mathchar'26\mkern-11mu{\rm d}}}                         %
\newcommand{\BEQ}{\begin{equation}}
\newcommand{\EEQ}{\end{equation}}
\newcommand{\BEA}{\begin{eqnarray}}
\newcommand{\EEA}{\end{eqnarray}}
\renewcommand{\H}{{\cal H}}
\renewcommand{\O}{{\cal O}}
\renewcommand{\d}{{\rm d }}

\newcommand{\e}{{\rm \varepsilon }}
\newcommand{\half}{\frac{1}{2}}

\newcommand{\Q}{\tilde Q}
\renewcommand{\S}{S_{\rm ep}}
\newcommand{\p}{\partial}
\newcommand{\minfty}{{-\infty}}

\newcommand{\I}{{\cal I}}

\newcommand{\nn}{\nonumber \\}
\begin{document}
\draft
\title{Thermodynamics of the glassy state:
effective temperature \\ as an additional system parameter}
\author{Th.~M.~Nieuwenhuizen}
\address{Van der Waals-Zeeman Instituut, University of Amsterdam
\\ Valckenierstraat 65, 1018 XE Amsterdam, The Netherlands}
\date{May 6, 1998; \today}
\maketitle
\begin{abstract}
A system is glassy when the observation time is much smaller than
the equilibration time.
A unifying thermodynamic picture of the glassy state is presented.
Slow configurational modes are in quasi-equilibrium
at an effective temperature. It enters thermodynamic relations
with the configurational entropy as conjugate variable.
Slow fluctuations contribute to susceptibilities
via quasi-equilibrium relations, while there is also a configurational term.
Fluctuation-dissipation relations also involve
the effective temperature.
Fluctuations in the energy are non-universal, however.
The picture is supported by analytically solving the dynamics
of a toy model.

\end{abstract}
\pacs{64.70.Pf, 75.10Nr,75.40Cx,75.50Lk}
\begin{multicols}{2}

Thermodynamics is an old but powerful subject.
After the invention of the steam machine
{it was needed for understanding their optimal efficiency.}
It applies to
a wide variety of systems, ranging from ideal gases to black holes.
Equilibrium thermodynamics
is a well understood subject. Important contributions were made by
Carnot, Clausius, Kelvin, Boltzmann, and Gibbs.

{Thermodynamics} for systems close to equilibrium was worked out
{in the mid} of this century. {Applications}
 are systems with
heat flows, electric currents, and chemical reactions. {A}
basic assumption is the existence of local thermodynamic equilibrium,
and the task is to calculate the entropy production.
Important contributions were made by de Donder, Prigogine, de Groot and
Mazur.

{Thermodynamics} for systems far from equilibrium has
long been a field of contradiction and confusion.
A typical application is a  window glass.
{A} cubic micron of glass is not a crystal, it is an undercooled liquid
which has fallen out of equilibrium.
In a glassy system the relaxation  time  $\tau_{eq}(T)$ of the slow
or so-called $\alpha$-processes has become
larger than the observation time, while the fast or $\beta$-processes
are still in equilibrium. Waiting long enough might bring the
system back to its equilibrium state.
In glasses one often assumes the Vogel-Tamman-Fulcher law
$\tau_{eq}\sim \exp(A/(T-T_K))$. In many other systems one encounters an
Arrhenius law $\tau_{eq}\sim\exp(A/T)$. As we shall demonstrate on
 a toy model, such systems display  the same glassy behavior.

We shall consider glassy transitions for liquids as well as for
random  magnets.
The results map onto each other by interchanging  volume $V$,
 pressure $p$, compressibility $\kappa=-\p \ln V/\p p$, and
expansivity $\alpha=\p \ln V/\p T$,
 by  magnetization $M$,  field $H$, susceptibility
$\chi=(1/N)\p M/\p H$,  and ``magnetizability'' $\alpha=(-1/N)\p M/\p T$,
respectively.

A glass forming liquid exhibits near the {glass}
transition smeared discontinuities in quantities such as the heat capacity
{$C_p$}, the expansivity and the compressibility. This defines a
smeared {glass transition} line {$T_g(p)$} or {$p_g(T)$},
with behavior similar to
continuous phase transitions of classical type, i.e., with specific heat
exponent $\alpha=0$. Since the twenties quite some attention has been
payed to introduce a thermodynamic description, see e.g.
{}~\cite{PdFbook}\cite{DaviesJones}\cite{Angell}.
It was investigated, in particular,
 whether the discontinuities satisfy
 the two Ehrenfest relations  (the analogs for second
order transitions of the Clausius-Clapeyron relation of a first order
transition), and whether the Prigogine-Defay ratio (see
eq. (\ref{PdF=})) {equals} unity.
Very recently we have explained the experimental observations~\cite{NEhren}.

A state that slowly relaxes to equilibrium is characterized by the
time elapsed so far, the ``age'' or ``waiting time''.
For glassy systems this is of special relevance.
In spin glass experiments  non-trivial
 cooling-heating cycles can be described by an effective age
 ~\cite{Hammann}. One thus
 characterizes a non-equilibrium state by three parameters,
$T$, $H$ or $p$, and the age $t$ or the cooling rate $\dot T=\d T/\d t$.
For  thermodynamics a more suitable third variable is
the {\it effective temperature} $T_e(t)$, introduced half a century
ago by Tool~\cite{Tool}~\cite{one}. $T_e\ne T$ will describe the
best quasi-equilibrium the $\alpha$-processes could reach so far;
it follows in simple cases  by equating $\tau_{eq}(T_e)=t$.

For a set of smoothly related cooling experiments $T_i(t)$
at fields $H_i$  one may express the effective
temperature as a continuous function: $T_{e,i}(t)$ $\to$ $T_e(T,H)$.
This sets a surface in $(T,T_e,H)$ space, that becomes multi-valued
if one first cools, and then heats. To map out the whole space
many sets of experiments are needed, e.g., at different cooling rates.
Thermodynamics amounts to give, for a certain class of  systems,
universal relations between state variables
at nearby points in this space.

Within this framework some new results were obtained
for a spin glass model with one step of replica symmetry breaking
{}~\cite{Nmaxmin}\cite{Nthermo}\cite{NEhren};
such models are related to systems without disorder that have one,
and only one, diverging time scale. The fast and slow modes do not
only have their own temperature, they also have their own entropy:
{the entropy of ``equilibrium processes'' $\S$,}
 and the ``configurational'' or ``information'' entropy
or ``complexity'' $\I$ {, respectively.}
The total entropy is $S=\S+\I$. {Previous}
 results can be summarized in {setting}
 $\dbarrm Q=T\d \S+T_e\d\I$. In the coarse of time  the system will
satisfy $(T-T_e)\d \I\ge 0$, as required by the second law.
In combination with the first law,
the thermodynamic relations can then be represented as
\BEA
\label{dU=}\d U&=&T\d \S+T_e \d \I+\dbarrm W\\
\label{F=} F&=&U-T\S-T_e \I\\
\label{dF=}\d F&=&-\S\d T-\I\d T_e+\dbarrm W
\EEA
where $\dbarrm W=-p\d V$ for liquids and $-M\d H$ for magnets.
Eq. (1) immediately leads to {the form}
$C_p=C_1+C_2\p T_e/\p T|_p$
{used in practice}
{}~\cite{Tool}\cite{DaviesJones}\cite{Petrosian}.

Out of equilibrium the Maxwell relation
between $U(T(t),t,H)\to U(T,T_e(T,H),H)$ and $M$
is, of course, not satisfied; however, the above implies that
its violation is related to the change of configurational entropy:
\BEQ\label{Maxwell}
\frac{\partial U}{\partial H}+M-T\frac{\partial M}{\partial T}=
(T_e-T\frac{\partial T_e}{\partial T} )\frac{\partial \I}{\partial H}
+T\frac{\partial T_e}{\partial H}\frac{\partial \I}{\partial T}
\EEQ

Here we wish to add to this scheme the non-equilibrium
susceptibilities. {We}
 will derive the results in a toy model with  fields $H_a$ and
 magnetizations $M_a$ {($a=1,2$)}.
As could {be}
 guessed from $M_a\equiv$
$M_a(T(t),T_e(t,H),H)$ with $T_e(t,H)\to T_e(T,H)$, there appear two terms:
\BEA\label{flucts=}
\chi_{ab}&\equiv&\frac{1}{N}\,
\frac{\partial M_a}{\partial H_b}\Bigl|_T\Bigr.
= \chi_{ab}^{\rm fluct}(t)+\chi_{ab}^{\rm conf}(t)
\EEA
First there is the expected fluctuation contribution
\BEA \label{chifluct}
\chi_{ab}^{\rm fluct}(t)&=&
\frac{ \langle \delta M_a(t)\delta M_b(t)\rangle_{\rm fast}}{NT(t)}+
\frac{\langle \delta M_a(t)\delta M_b(t)\rangle_{\rm slow}}{NT_e(t)}\nn
&-&\frac{1}{N}\frac{\partial M_a}{\partial T_e}\Big|_{T,H}\,\,
\frac{\p T_e}{\partial H_b}\Bigr|_{t}
\EEA
{Notice that}
fast and slow processes again enter with their own temperature,
{while  the third term is small.} {Since}
$T_e\neq T$, {there occurs}
a new, configurational term
\BEA \label{chiconf}
\chi_{ab}^{\rm conf}=\frac{1}{N}\,
\frac{\partial M_a}{\partial T_e}\Bigl|_{T,H}\,\,
\frac{\partial T_e}{\partial H_b}\Bigr|_{T}
\EEA
It originates from the difference in  the system's
structure for cooling experiments at nearby fields ~\cite{GoldsteinJaeckle}.
Such universal relations would not hold for energy fluctuations.

We shall also see in the toy model, that the
correlation function $C_{ab}(t,t')$$=$$(1/N)
\langle \delta M_a(t)\delta M_b(t')\rangle$ and the response function
$G_{ab}(t,t')$$=$$(1/N)\delta M_a(t)/\delta H^b(t')$
satisfy the fluctuation-dissipation relation
\BEQ \label{FDR=}\frac{\partial C_{ab}(t,t')}{\p t'}
=T_e(t'){G_{ab}(t,t')}\EEQ
in the aging regime, while in the equilibrium or
short-time regime $T$, replaces $T_e$. This
been confirmed numerically for a soft sphere glass~\cite{Parisi}.
Eq. (\ref{FDR=}) is consistent with
(\ref{chifluct}): in performing $\int_0^t G_{ab}(t,t')\d t'$
one uses that $\p_ {t'} C$ changes much faster than $T_e(t')$,
allowing to
replace $T_e(t')$ by $T_e(t)$ in the aging regime, and by $T$ in the
equilibrium regime $t'\approx t$.

Note that the ratio $\p_{t'}C/G$ depends on time.
The situation with constant $T_e$ is well known from mean
field spin glasses~\cite{BCKM}.
However, this time-independence is an artifact
of the mean field approximation~\cite{Nthermo}.

Finally, one estimates scaling in the aging regime of two-time
quantities as
\BEQ
C(t,t')\approx C(\frac{t-t'}{\tau_{eq}(T_e(t'))})
\approx C(\frac{t-t'}{t'})=C(\frac{t}{t'})\EEQ
showing immediately the familiar $t/t'$ scaling; there may be
logarithmic  scaling corrections.

The full  picture (eqs. (1)-(9))  says that
slow modes are at a quasi-equilibrium at
 $T_e$. As in a plasma, slow and fast modes equilibrate
at their own temperature.

In a number of simple models there occurs a dynamical glassy state
when cooling near $T=0$. If there is only one time-scale,
 there remain only slow processes in the frozen phase
 (only $\alpha$-, no $\beta$-processes). This implies that $U$
has no explicit dependence on $T$, and that $\S=0$.

In order to corroborate our statements, we consider a toy model
involving free spherical spins
{ ($\sum S_i^2=N$).}
 The Hamiltonian contains two parts,
a ``self-interaction'' term involving quenched
random fields $\Gamma_i=\pm\Gamma$ with average zero,
and a coupling to an external field $H$
\BEQ \H=-\sum_{i=1}^N \Gamma_i S_i-H\sum_{i=1}^N S_i\EEQ
In terms of the ``staggered''
magnetization $M_s\equiv (1/\Gamma)\sum_i \Gamma_i S_i$
one simply has $\H=-\Gamma M_s-HM$.

In equilibrium at low $T$ the internal energy reads
$U_{eq}/N=-K+\half T$, the entropy $S_{eq}/N=
\half\ln(eT/K)$, the magnetizations
$M_{eq}/N=H/K-\half HT/K^{2}$, and
 $M_{s;\,eq}/N=\Gamma/K-\half \Gamma T/ K^{2}$,
where $K\equiv \sqrt{\Gamma^2+H^2}$.

This model gets glassy behavior when it is subject to
Monte Carlo dynamics with parallel update, which couples the spins
dynamically. The time evolution can be solved exactly,
since it maps closely on the dynamics
for uncoupled harmonic oscillators, introduced
by Bonilla, Padilla, and Ritort~\cite{BPR},
when extended to include  a field.

Per time step $1/N$ one makes  parallel Monte Carlo moves,
$S_i\to S_i'=S_i+r_i/\sqrt{N}$, with
Gaussian noise having
$<\!\!r_i\!\!>=0$ and $<\!\!r_i^2\!\!>=\Delta^2$.
 Next one makes a global rescaling of the length of the spins in order
to keep $\sum S_i'^{\,2}=N$.
This leads to the  final update
\BEQ
S_i'=S_i+\frac{r_i}{\sqrt{N}}-S_i\sum_j(\frac{r_jS_j}{N\sqrt{N}}
+\frac{r_j^2}{2N^2})+\cdots
\EEQ
Denoting $\H$ by $E$, it is simple to calculate the joint transition
probability of $x\equiv E'-E$ and $y\equiv M'-M$:
\BEQ p(x,y|E,M)=
p(x|E)p(y|x,E,M) \EEQ
where both factors are Gaussian.
In terms of
\BEQ \e=K+\frac{E}{N}; \quad m=\frac{M}{N};\quad
\tilde m =m-\frac{H}{K}+\frac{H}{K^2}\e
\EEQ
the centers of the Gaussians are
\BEA
&x_0 =\half\Delta^2(K-\e); \quad
y_0=-\frac{Hx}{K^2}-\tilde m\frac{\Delta^2K^2-2Kx+2x\e}{2\e(2K-\e)}
\EEA
while  their variances read
\BEQ
\Delta_x=\Delta^2\e(2K-\e)  ;\quad
\Delta_y=\frac{\Delta^2\Gamma^2}{K^2}
-\frac{\Delta^2K^2 \tilde m^2}{\e(2K-\e)},
\EEQ
respectively. Following Metropolis,
a move $E'=E+x$, $M'=M+y$ is accepted with probability
$W(\beta x)$, where $W=1$ if $x<0$, while $W=\exp(-\beta x)$ if $x>0$.
The dynamics is now fully specified. The average energy evolves
according to a closed equation for $\e(t)$ ~\cite{BPR},
\BEQ\label{dedt=}
\frac{\d \e}{\d t}=\int_\minfty^\infty \d x W(\beta x)x p(x|\e)
\EEQ
We find that the average magnetization satisfies similarly
\BEQ
\frac{\d m}{\d t}=\int_\minfty^\infty \d x W(\beta x)y_0 p(x|\e)
\EEQ
At low $T$ there occur Arrhenius laws for the
equilibrium relaxation times
viz. $\tau_{eq}\equiv \tau_{eq}^{(E)}=\half\beta A\tau_{eq}^{(M)}=
(\pi T/64 A)^{1/2} \exp(A/T)$ with $A\equiv\Delta^2K/8$, responsible
for non-equilibrium behavior in typical cooling procedures.

We first consider the situation at $T=0$. Starting from a random initial
condition, the system will slowly evolve towards the ground state.
It is this evolution that we now wish to capture within a thermodynamic
framework. Equating $U(t)=U_{eq}(T_e,H)$
leads to~\cite{BPR}
\BEQ \label{Tet=}
T_e(t)\approx \frac{A}{\ln 2t/\sqrt{\pi} +\ln\ln 2t/\sqrt{\pi}}
\EEQ
As asserted above, the same $T_e$ is obtained via the relaxation time:
$t=\tau_{eq}(\tilde T_e(t))$  $\to$
$ \tilde T_{e}(t)=T_e(t)+\O(T_e^2)$.

In equilibrium it holds that $\tilde m=0$. The result
 $\tilde m \sim 1/t\sim \exp(-\beta_e A)$
proves that the magnetization very closely follows
its quasi-equilibrium  value $m_{eq}(T_e,H)$.

To test the thermodynamics we need the configurational entropy.
It is defined as the logarithm of the number of states leading
to $U(t,H)=U_{eq}(T_e,H)$. Since $\S=0$ it simply holds that
$\I=S_{eq}(T_e(t),H)$. We can now verify the relation
$\d U=T_e \d \I-M\d H-M_s\d\Gamma$.
At constant $H$ and $\Gamma$ it is valid,
because one has replaced $T\to T_e$ in energy and entropy.
Then one can take the difference of two evolution experiments
at two nearby $H$'s. The relation remains satisfied since
$M(t)\approx M_{eq}(T_e,H)$.
The modified Maxwell relation (\ref{Maxwell}) is also obeyed.
Finally one can change $H$ in the coarse of time.
Then $\tilde m\ll HT_e/K^2$ as long
$\partial H(t)/\partial T_e(t)\ll \beta_e(t) H(t) \Delta^2$.
{This} is a mild condition.
If $H$ is changed quicker, or if it goes to zero too rapidly,
the system will not be able to reach a quasi-equilibrium
{described by $T_e$ alone}.

We have also considered the fluctuation formula. They are
too lengthy to be reproduced here. We found
\BEA
\frac{\langle \delta M^2\rangle}{N}&=&\frac{\Gamma^2 T_e}{K^3}
{-\frac{\Gamma^2T_e^2}{2K^4}+ {\cal O}(T_e^3)} \nonumber\\
\frac{\langle \delta \H^2\rangle}{N}
&=&-\frac{K^2\langle \delta \H\delta M\rangle}{HN}=\frac{T_e^3}{A}
\EEA
The latter two relations imply that
$\Gamma^2\langle \delta M_s^2\rangle$
$\approx$ $H^2\langle \delta M_s^2\rangle$
$\approx$ $-\Gamma H\langle \delta M\delta M_s\rangle$.
Since there are no fast processes, it immediately follows that the
quasi-equilibrium relation (\ref{flucts=})
is satisfied for all four cases, to leading order in $T_e$.
In contrast, for $M_a\to \H$, $H_a\to \beta$ or $\beta_e$
the corresponding relations are violated.

The two-time correlation functions
$C_{ab}(t,t')$ and response functions  $G_{ab}(t,t')$
exhibit related behavior.
These functions decay as $C_{ab}(t,t')=C_{ab}(t',t')h(t')/h(t)$
with $h(t)=\exp(\beta_e(t)A)\sim t\sqrt{\ln t}$ and satisfy
the fluctuation-dissipation relation (\ref{FDR=})
for any non-zero value of $H$ and $\Gamma$~\cite{notice}.
(For the oscillator model at zero field these relations were derived
in ~\cite{BPR}.)

Now we consider a cooling experiment from high temperatures.
A {glass} transition will occur when the
{cooling  timescale $T/|\dot T|$}
becomes comparable to
{ the equilibration timescale $\tau_{eq}(T_e(t))$.}
Let us assume that it happens at a low temperature
$T_ {g} \ll A$,
so at an exponential time scale $t_{g}
=\tau_{eq}(T_{g})\sim \exp\beta_{g} A$.
This will imply that the width of the
{transition} region is small: $\Delta T_ {g}\sim T_{g}^2/A\ll T_{g}$. 
Assuming to have in that region
a constant dimensionless cooling rate $\Q=\dot T\tau_{eq}'(T)$,
we derive from (\ref{dedt=})
\BEQ \label{dTedT}
 \frac{\p T_e}{\p T}=\frac{T-T_e}{\Q}\,\,
\frac{\tau_{eq}'(T_{g}+(T-T_{g})/\Q)}{\tau_{eq}(T_e)}\EEQ
This equation is similar to but different from
earlier proposals~\cite{Tool}~\cite{Petrosian},
and might be universal for narrow glassy transitions.
It covers three cases: a) $\Q>1$: normal cooling
 towards or in the glass;
b) $0<\Q<1$: cooling in
a glassy state so slowly that equilibrium is achieved later;
c) $\Q<0$:  heating up in the glassy state.
The apparent  specific heat $c=\half \p T_e/\p T$
decreases {if}
 cooling is not too slow. In a heating experiment
it is negative, but it produces near $T_{g}$ the well know overshoot
$c>\half$,  with height and shape depending solely on $\Q$.

In a cooling experiment from large $T$
eq. (\ref{dTedT}) yields initially
$T_e(t)=T(t)$, up to exponential corrections,
describing thermodynamic equilibrium at the instantaneous temperature.
Below the {glass transition} region one has
$T_e\approx T_{g}+(T-T_g)/\Q$  $\approx$
$\tau_{eq}^{-1}(t)$. This agrees with  (\ref{Tet=}) and
shows that the actual temperature and the cooling history
are irrelevant: to leading order the energy just evolves as
if the system had been quenched to zero temperature, and aged there.
It can be checked that all relations linear in $T_e$
remain as at $T=0$, thus supporting the picture (1)-(9)
{}.

This solution allows us to check the Ehrenfest relations
along the {glass transition} line $H_{g}(T)$ 
in cooling procedures ($\Q>1$). Below the {transition}
region  one has $c=1/(2\Q)$,
$\alpha =H/(2K^2\Q)$
and  $\chi=\Gamma^2/K^3 -(H/2K^2)$ $\partial T_e/\partial H$.
Comparing with the paramagnet
and using that along the {transition}
line the equality $T_e=T$
 implies that
$\partial T_e/\partial T+(\partial T_e/\partial H)(\d H_{g}
/\d T)=1$,
we find that the jumps in $\alpha$ and $\chi$ satisfy
the first Ehrenfest relation
\BEQ \label{Ehren1}
\Delta \alpha=\Delta \chi\,\frac{\d H_{g}
}{\d T} \EEQ
Until very recently, it was widely believed that this
 relation is violated at the glass transition.
We pointed out  that experimentalists had
inserted some short-time value of $\chi$ or $\kappa$~\cite{NEhren},
like the ``zero-field-cooled'' susceptibility $\chi_{ZFC}$,
i.e. the first term of $\chi^{\rm fluct}$ in eq. (\ref{chifluct}).
 In spin glasses with one step replica symmetry breaking
$\chi_{ZFC}$ is strictly smaller than the $\chi$ of the paramagnet
{}~\cite{pspin}.
In the present model, there are no $\beta$-processes, so
$\chi_{ZFC}$ even vanishes. On the other hand, the long-time or
field-cooled value $\chi_{FC}$$=$$\chi^{\rm fluct}$ is continuous at
$T_{g}$.
The correct discontinuity, $\Delta\chi=-\chi^{\rm conf}$
arises from eq. (\ref{chiconf}).
As $\Delta\alpha=(\p m/\p T_e)(\p T_e/\p T-1)$,
this explains in detail  why the first Ehrenfest
relation is satisfied automatically~\cite{NEhren}.

It was also shown in ~\cite{NEhren} that  the modified
Maxwell relation (\ref{Maxwell}) leads to
the modified second Ehrenfest relation
\BEQ
\frac{\Delta C}{NT{_g}}=\Delta \alpha\frac{\d H}{\d T}+
\frac{1}{N}(1-\frac{\partial T_e}{\partial T})
(\frac {\p \I}{\p T}+\frac {\d H_{g}}{\d T}\,\frac {\p \I}{\p H})
\EEQ
We can now verify that it is also satisfied.
The new last term needed for validity beyond equilibrium.

 The Prigogine-Defay ratio may be expressed as
\BEQ\label{PdF=}
\Pi\equiv \frac{\Delta C\Delta\chi}{NT_{g}(\Delta \alpha)^2}
=1+\frac{1}{N\Delta \alpha }(1-\frac{\partial T_e}{\partial
T}) \frac {\d \I}{\d H} \EEQ
The definition looks as an equilibrium relation,
and it shown that $\Pi$ must be larger than unity
for mechanical stability~\cite{DaviesJones}.
This was based, however, on the invalid assumption of
thermodynamic freezing of a set of unspecified order parameters.
The  equivalent relation
$\Pi=(\Delta C/NT_{g} \Delta \alpha )\d T_{g}
/\d H$ allows $\Pi<1$
if $\d T_{g}/\d H$, which depends on $\d Q/\d H$, is small enough.
We realized that already in the  classic experiment
of  Rehage and Oels~\cite{RehageOels} on the glass transition
in atactic polystyrene there occurs a value $\Pi=0.77$~\cite{NEhren}.
In our present model $\Pi=(K^2/HT_{g})\d T_{g}/\d H$
becomes less than unity whenever $\d \lambda_{g}/\d H$ is positive.
This condition  occurs in half of the
smoothly related sets of cooling trajectories.

In conclusion, we have proposed a unifying thermodynamic
picture of the glassy state.
It does not apply to ideally slow experiments, but to
conditions that are typically met.
As in a plasma, slow and fast modes equilibrate
at their own temperature.
 On long time scales and under mild conditions,
global thermodynamical quantities,
like the energy and volume or magnetization(s),
go to quasi-equilibrium values at a certain
effective temperature $T_e(t)$. Slow fluctuations contribute to
susceptibilities with factor $1/T_e$.
For cooling trajectories at two nearby external fields (pressures),
the difference of the magnetization (volume)
involves the usual fluctuation susceptibility (compressibility),
and a new structural contribution, that arises  from the difference in
effective temperatures.
Correlation and response functions exhibit $t/t'$ scaling,
and satisfy a  fluctuation-dissipation  relation involving  $T_e$.

We have verified our picture in
the glassy phase of the low-temperature dynamics of an
exactly solvable toy model, that contains two external fields.
We expect that the fields may also stand for other ``mechanical''
forces, such as pressure or chemical potential.

The picture has the right signs to be valid for
a class of glassy systems.
For verifying it in glass forming liquids it is desirable to map
out the full ($T,T_e,p$)-space by cooling experiments,
and to check that against aging experiments.


The author thanks J. Groeneveld, J. J\"ackle,
P.G. Padilla, F. Ritort, and M. Sellitto
for discussion, and the Newton Institute (Cambridge, UK) for hospitality.


\end{multicols}


\references
\bibitem{PdFbook} I. Prigogine and R. Defay,
{\it Chemical Thermodynamics},
(Longmans, Green and Co, New York, 1954)

\bibitem{DaviesJones} R.O. Davies and G.O. Jones,
 Adv. Phys. {\bf 2} (1953) 370

\bibitem{Angell} C.A. Angell, Science {\bf 267} (1995) 1924

\bibitem{NEhren} Th.M. Nieuwenhuizen, Phys. Rev. Lett. {\bf 79}(1997) 1317

\bibitem{Hammann} F. Lefloch, J. Hammann, M. Ocio, and E. Vincent,
Europhys. Lett. {\bf 18} (1992) 647

\bibitem{Tool} A.Q. Tool, J. Am. Ceram. Soc. {\bf 29} (1946) 240.

\bibitem{one}
{It has been}
supposed that also an effective pressure
or field should be introduced. It is not needed here.

\bibitem{Nmaxmin} Th.M. Nieuwenhuizen,
 Phys. Rev. Lett. {\bf 74} (1995) 3463; cond-mat/9504059

\bibitem{Nthermo} Th.M. Nieuwenhuizen, J. Phys. A {\bf 31}
(1998) L201

\bibitem{Petrosian} V.P. Petrosian,
Ru. J. Phys. Chem. {\bf 69} (1995) 183




\bibitem{GoldsteinJaeckle}
Such a term was anticipated.
M. Goldstein (J. Phys. Chem. {\bf 77} (1973) 667) notices that
$V_{glass}$ depends stronger on the pressure of formation $\hat p$ than on
the one remaining after partial release of pressure.
J. J\"ackle (J. Phys: Condens. Matter {\bf 1} (1989) 267; eq, (9))
then assumes that for infinitely slow cooling $\hat p$
is the only additional system parameter, and argues that
$\Delta \kappa_T\to\Delta\kappa= \Delta \kappa_T
+\partial \ln V/\partial \hat p =\Delta\alpha\,\d T_{g} /\d p$
and  $\Pi=\Delta \kappa_T/\Delta \kappa>1$.

\bibitem{Parisi} G. Parisi,
Phys. Rev. Lett. {\bf 79} (1997) 3660


\bibitem{BCKM}
J.P. Bouchaud, L. F. Cugliandolo, J. Kurchan, and M. M\'ezard,
Physica A {\bf 226} (1996) 243 review that in mean field spin glasses
the fluctuation-dissipation
ratio $X(t,t')$$\equiv$$ TG(t,t')/\partial _{t'}C(t,t')$ simplifies to
$X(t,t')\equiv \hat X(C(t,t'),t')$$=\hat X(0,t')\to
const$. As $T_e(t')=T/\hat X(0,t')$ governes our physics, the
$t'$-dependence of our $\hat X$ cannot be omitted.
Only at exponential time scales the mean field spin glass
is related to realistic systems ~\cite{Nthermo}.

\bibitem{BPR} L.L. Bonilla, F.G. Padilla, and F. Ritort,
Physica A, to appear; preprint cond-mat/9706303

\bibitem{notice}
{If $\Gamma$ or $H$ vanishes, the model becomes too poor.}


\bibitem{pspin}
At $H=0$ one has $\chi_{ZFC}=\beta(1-q_{EA})$, while $\chi_{FC}=
\beta(1-(1-x_1)q_{EA})$ matches $\chi_{PM}=\beta$ at $x_1=1$.


\bibitem{RehageOels} G. Rehage and H.J. Oels,
High Temperatures-High Pressures {\bf 9} (1977) 545
\end{document}